\documentstyle[emulateapj,onecolfloat,psfig]{article}

\newcommand{\lin}{{\rm lin}}

\newlength{\tskip}
\newlength{\colwidth}

\newcommand{\beq}{\begin{equation}}
\newcommand{\eeq}{\end{equation}}
\newcommand{\beqa}{\begin{eqnarray}}
\newcommand{\eeqa}{\end{eqnarray}}

\begin{document}
\twocolumn[
\title{Are galaxy properties only determined by the dark matter halo mass?}
\author{Asantha Cooray}
\affil{Department of Physics and Astronomy, University of California, Irvine, CA 92697}

\begin{abstract}
Conditional luminosity function (CLF) of dark matter halos supersedes simple models
on the number of galaxies as a function of the halo mass, or the so-called
halo occupation number, by assigning a luminosity distribution to those galaxies. 
Suggestions have now been made that both the clustering strength and properties of hosted galaxies  
of a given dark matter halo depends also on the age and the environment of that halo in addition to halo mass. 
Based on simple CLF models, we find that at least one suggestion that made use of
central galaxy properties may be affected by uncertainties in the group catalog from which 
clustering properties were measured to address the age dependence. Establishing
how galaxy properties  of fixed mass halos change with the environment or the age 
is challenging, if not impossible, given uncertainties in group and cluster catalogs.
It may be possible to address this issue through
statistics based on a combined study of luminosity and color distributions, and
luminosity- and color-dependent galaxy clustering, all of them as a function of the galaxy overdensity,
though no single statistic is likely to provide the ultimate answer.
The suggestion that the age dependence of the halo bias invalidates the analytical 
models of the galaxy distribution is premature given the improvements associated with 
modeling approaches based on conditional functions.
\end{abstract}

]

\section{Introduction}
Recently, several claims have been made that the galaxy properties of a given dark matter halo
may depend on the environment in which the halo resides (Gao et al. 2005; Harker et al. 2005; Yang et al. 2005b). 
A connection is generally made to the age of the halo since age may be the primary factor that 
determines the environment in which the halo forms. Numerical simulations of the dark matter
distribution suggests that older halos are more strongly clustered than younger halos of a given mass (Gao et al. 2005). 
Such an age dependence on the halo bias is not yet present within successful analytical descriptions of the halo
distribution and clustering properties (e.g., Mo \& White 1996; Sheth, Mo \& Tormen 2001).
These results invalidate analytical models of the galaxy distribution 
since these models assume that the galaxy properties are
only determined by the halo mass (Kravtsov et al. 2004; Cooray \& Sheth 2002 for a review).

Beyond simulations, whether galaxy properties of a given halo mass vary
with the  environment is yet to be established observationally.
The galaxy morphology--density (Dressler 1980; Dressler et al. 1997) and
the color--density (Blanton et al. 2004) relations do indicate that
galaxy properties change with the environment: red, early-type, galaxies are usually found in denser environments than the 
environments where blue, late-type, galaxies are found. 
It could, however, be that the increasing fraction of early-type galaxies with increasing density is simply a reflection of a higher
fraction of early-type galaxies in more massive halos (e.g., Cooray 2005a; Weinmann et al. 2005).  
Thus, interpretations of the density-morphology or density-color relations are generally inconclusive with some suggesting support
for the halo model (Blanton et al. 2004) while others not (Goto et al. 2003; Balogh et al. 2004).

To study the difference between galaxy properties of same mass halos when either the environment or the age 
is varied one must find fixed mass halos while a second variable, such as the environment, is varied.
The test related to the environment could be 
achieved with large catalogs of galaxy groups, but since one may not be able to reliably assign a mass 
to an individual group, it is unlikely that there is an easy way to uniquely identify whether one is comparing
halos of similar mass. Since the expected differences are likely to be at the few to, at most, ten percent level (see
e.g., discussion in Zheng \& Weinberg 2005), the mass assignments must be accurate to the same precision.
Statistics that measure properties after randomizing galaxies 
can partly address the question on the environmental dependence by
testing whether the surrounding large-scale environment is responsible for properties within a halo. It could be that the
internal structure of a halo, such as the dark matter concentration parameter, 
changes with the environment and galaxy properties reflect those changes.

The second approach to address this question involves statistical study of a large sample of galaxies, 
such as clustering, binned in terms of an estimator of age for galaxies.
Using a catalog of galaxy groups and clusters from the 
2-degree Field Galaxy Redshift Survey  (2dFGRS; Colless et al. 2001),  
Yang et al. (2005b) recently considered the cross-correlation function
between the central bright galaxies in galaxy groups and the general population of galaxies above a certain
luminosity.  For groups with halo masses above $M_\star$, they found the groups with a lower luminosity central galaxy to be
more strongly clustered than the ones with a higher luminosity galaxy. 
Furthermore, galaxies that are more passive in a given mass bin were
found to be more correlated. Thus, a claim was made that the cross-clustering measurement in
Yang et al. (2005b) indicates age and environmental dependence of halo bias.

Using models of the conditional luminosity function (CLF; Yang et al. 2003, 2005a; Cooray \& Milosavljevi\'c 2005b),
we make a simple model of the cross-correlation between central galaxies of groups and the 
galaxy distribution and suggest that the measurement in Yang et al. (2005b) may be contaminated by 
uncertainties in their group catalog involving mass assignments.
We find that, for example,  if one selects roughly 10\% to 20\% fraction of satellite galaxies of larger mass halos
mistakenly as the central galaxy in lower mass halos, then the bias factor for smaller halos with fainter luminosities
is boosted.  With a simple relation between the starformation rate and
the galaxy luminosity, the clustering results binned in terms of starformation rate may
also be explained without requiring a second parameter.
It is unlikely that Yang et al. (2005b) measurements provide a definitive answer 
on the age dependence of galaxy properties of fixed halo mass with age captured by
the either the star-formation rate or the luminosity of the central galaxy. 

Here, we also consider if there is an appropriate statistical test on the density and age dependence of galaxy properties 
within a given halo that is affected less by systematics.
While we have yet to come up with a simple test to provide the ultimate answer
a combination of measurements on the luminosity and color distributions, combined with clustering,
all as a function of galaxy density may be utilized to address this issue.
Any single statistical measurement is affected by two potential variations: the mass function that varies
the halo distribution with the environment and galaxy properties within a fixed halo mass that change with environment:
\begin{equation}
P(L,\delta_{\rm gal})  \propto \int dM\; \frac{dn(\delta_{\rm gal})}{dM} \Phi(L|M,\delta_{\rm gal})
\end{equation}
where both the mass function,  $dn/dM$, and the CLF, $\Phi(L|M)$, varies
 as a function of the environment captured by $\delta_{\rm gal}$, which is defined in terms 
of the density of galaxies in some fixed volume.  To address the question whether
 $\Phi(L|M,\delta_{\rm gal})=\Phi(L|M)$ observationally, the statistical methods must allow a separation of effects coming from 
$dn/dM(\delta_{\rm gal})$ with $\Phi(L|M,\delta_{\rm gal})$ (see, e.g., Mo et al. 2004; Cooray 2005a).
This degeneracy is also what affects interpretation of the density-morphology or density-color relations.
 
The {\it Letter} is organized as following: In the next section, we briefly summarize a model for
the cross-clustering between central galaxies in galaxy groups and the general galaxy distribution.
 We compare our model predictions with those of Yang et al. (2005b) and suggest that
their observations can be explained in terms of a small contamination in the group catalog. Furthermore, we comment
on potential tests to address whether  $\Phi(L|M,\delta_{\rm gal})=\Phi(L|M)$ or not.
Finally, as a side note, even if it turns out that the CLF has a second parameter, we will argue that
this does not imply that the analytical halo model is wrong, but rather it can be easily improved by including
the environmental   dependence when describing galaxy properties through conditional functions.

\begin{figure*}[!t]
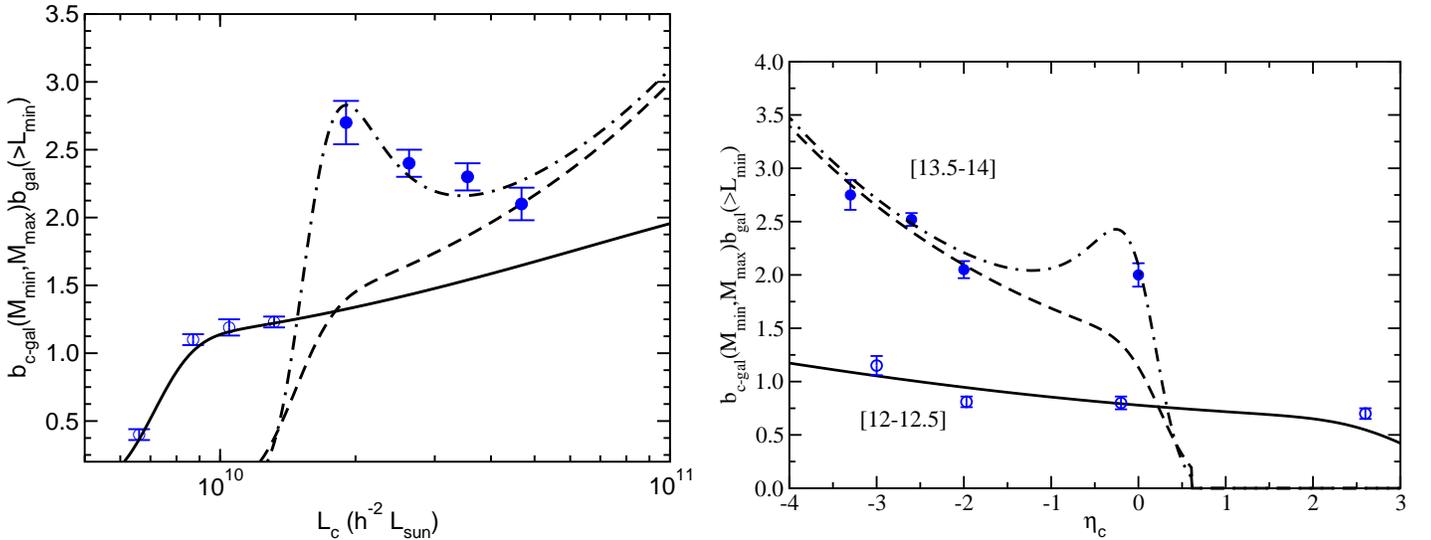

\centerline{\psfig{file=bosch.eps,width=3.6in,angle=0}
\hspace{0.1cm}
\psfig{file=biaseta.eps,width=3.6in,angle=0}}
\vspace{0.3cm}
\caption{The product of bias factor  of central galaxies in groups $b_c(L_c)$ and
galaxy bias for a volume limited sample above a $b_J$ magnitude of $-19.45+5\log h$  
with measurements subdivided to mass bins in galaxy groups that host the central galaxy.
The measurements are from Yang et al. (2005b) and we show their data for
two mass bins from  10$^{12}$ to $10^{12.5}$ (open symbols) and $10^{13.5}$ to 10$^{14}$ M$_{\sun}$ (filled symbols).
In the left panel we consider measurements as a function of central galaxy luminosity while in the right panel we 
consider their measurements as a function of star-formation rate.
The solid and long-dashed lines show the expected bias factor for central galaxies in these mass bins as
a function of the central galaxy luminosity.
The dot-dashed line shows a simple modification to the bias factor calculation where we include
a small correction to central galaxy sample by allowing satellites in larger mass halos ($> 10^{14}$ M$_{\sun}$)
to be identified (wrongly) as central galaxies  and be assigned to a group with a mass in the range
between $10^{13.5}$ M$_{\sun}$ to 10$^{14}$ M$_{\sun}$. The contaminant fraction $f$ (see text) needed to explain
left and right panels differ by a factor of 2 and could indicate extra scatter in mapping between
luminosity, which is the primary ingredient in the CLF-based models, to the star-formation rate of central galaxies.
The simple analytical description, which does not make any modifications to the halo bias 
as a function of age, suggests that these measurements with the observed galaxy distribution do not conclusively 
establish an age dependence to galaxy clustering properties.}
\label{fig:sloan}
\end{figure*}

\section{Galaxy bias of Central Galaxies in Groups}

In Yang et al. (2005b), the cross-correlation between luminous galaxies (presumed to be at halo centers)  and a large sample of
galaxies, above some luminosity, is studied as a function of the luminosity and star-formation rate of the central galaxy.
In models based on the conditional luminosity function, the bias factor of galaxies above a minimum luminosity $L_{\rm min}$ 
can be calculated as
\begin{eqnarray}
&& b_g(> L_{\rm min}) = \\
&& \left[\int_{L_{\rm min}} dL\;\Phi(L)\right]^{-1} \int_{L_{\rm min}} dL\int dM\; \frac{dn}{dM}\; b_{\rm halo}(M)\; \Phi(L|M) \, , \nonumber
\end{eqnarray}
where the galaxy LF is $\Phi(L)=\int dM\; dn/dM\; \Phi(L|M)$ and $b_{\rm halo}(M)$ is the halo bias.
We make use of Sheth \& Tormen (1999) description for the mass function and the halo bias, consistent with numerical simulations for both
the halo mass (Jenkins et al. 2001) and average bias (Seljak \& Warren 2004).
This CLF $\Phi(L|M)$ can be divided to central and satellite galaxies following the
approach advocated by Cooray \& Milosavljevi\'c (2005b). The central galaxy luminosities are assigned to halo masses
through the Cooray \& Milosavljevi\'c (2005a) relation as appropriate for the 2dF $b_J$ band 
(Cooray 2005a; also, Vale \& Ostriker 2004) including a scatter with a log-normal dispersion of 0.17
as measured from SDSS LF and clustering  data (Cooray 2005b).

We write the central galaxy bias when divided to  mass bins as a function of the central galaxy luminosity or star-formation rate $\eta$  as
\begin{eqnarray}
&&b_c(L \; {\rm or} \;  \eta|M_{\rm min},M_{\rm max}) = \\
&& \frac{\int dM\; dn/dM\; b_{\rm halo}(M)\; \left[h(M,M_{\rm min},M_{\rm max})\Phi_c(L|M)\right]}{
\int dM\; dn/dM\; \left[h(M,M_{\rm min},M_{\rm max})\Phi_c(L|M)\right]} \, , \nonumber
\end{eqnarray}
where $h(M,M_{\rm min},M_{\rm max})$ is a function such that $h(M,M_{\rm min}$ 
$M_{\rm max}) =1$
when $ M_{\rm min} \lesssim M \lesssim M_{\rm max}$.  While 
the group catalog is divided to bins in estimated halo masses,
we allow for an uncertainty in both the minimum and maximum mass of each bin
since it could be that halos with true masses below and above 
are assigned incorrect masses within the bin through statistical uncertainties.
We achieve this with an analytical function that is essentially flat across the mass range with tails both above and below the mass
cut-offs, and normalized to be unity. In  addition to mass limits,  we also allow for a slight variation in the luminosity cut-off.
In Yang et al. (2005b), the galaxy sample used involve a volume limited sample with galaxies $M < -19.45 + 5 \log h$ in the $b_J$-band.
In our CLF models, instead of a sharp cut-off at $L_{\rm min}$ corresponding to this absolute magnitude cut, we allow for a slight scatter in
this luminosity as well.

The predicted large-scale clustering is shown in Fig.~1. Here, we plot the product of $b_g(>L_{\rm min})b_c(L_c|M_{\rm min},M_{\rm max})$
as a function of $L_c$, the luminosity of central galaxies, for mass bins defined by $M_{\rm min} < M < M_{\rm max}$.
Instead of all mass bins considered in Yang et al. (2005b), we consider two bins at the low and high mass end. 
The solid and long-dashed lines show the expected bias factor for two bins between 10$^{12}$ to $10^{12.5}$ and
$10^{13.5}$ to 10$^{14}$ M$_{\sun}$. When compared to the measurements from Yang et al. (2005b), CLF models based on
standard description for halo bias and mass function show  a lower bias factor 
for low luminosity galaxies assigned to higher mass groups.
Similarly, we also consider measurements divided based on star-formation rate ($\eta$). Since our model is based on luminosities we need
a mapping between $L_c$ and $\eta_c$. We considered a simple one to relation of the form $\eta_c=\alpha(L_c/10^{10} \; L_{\sun})+\beta$
and found out that combinations of $(-5,1.5)$ and $(-8,3.5)$ describe the data best in the high- and low-mass bins respectively.
The difference in the mapping between $L_c$ and $\eta_c$ for the two mass bins suggest that starformation rate of a central galaxy
depends on the mass of the halo in which it resides. 

To study the bias factor in the case of contaminations, we modify 
square brackets in Eq.~3 as $[h(M,M_{\rm min},M_{\rm max})$ $\Phi_c(L|M)]$ $\rightarrow$
$[h(M,M_{\rm min},M_{\rm max})$ $\Phi_c(L|M)$ $+ f \Phi_s(L|M)$ $\Theta(M,M_1,M_2)]$, where $\Theta(M,M_1,M_2)$ is the unit step function such
that either $\Theta(M,M_1,M_2) =1$ when $M_1>M>M_2$ or zero otherwise.  The assumption here is that
some fraction of central galaxies are satellites of larger mass halos (above the mass range corresponding to a given bin),
but mistakenly taken to be central galaxies by the group finding algorithm, say, 
and assigned with a lower halo mass.  
With $M_1=10^{14}$ and  $M_2=10^{14.5}$, $f \sim 0.002$ leads to the dot-dashed line shown in Figure~1(right panel),
while a value of $f\sim 0.001$ describes the dot-dashed line in Figure~1(left panel) and with $L_c$ to $\eta_c$
mapping described above. The difference in $f$ between the two plots may be suggestive of
extra scatter in the mapping between $L_c$ and $\eta_c$.
Note that $f$ should not be interpreted simply as the number of central galaxies that may be contaminated in the group
catalog  since in our model description, a large difference exists
between CLFs for central and satellite galaxies. Integrating over the LF, the difference in terms of a fraction  of galaxies is at
the level of 12\% percent in the lowest luminosity end of the mass bin between 10$^{13.5}$ M$_{\sun}$ to 10$^{14}$ M$_{\sun}$
but drops to the level of a few percent and below as the luminosity is increased; this luminosity-dependent change in the contamination is
not a reflection of a complex situation but rather the fact that in the mass bin  10$^{13.5}$ M$_{\sun}$ to 10$^{14}$ M$_{\sun}$
there are more ``true'' central galaxies at the high luminosity end.
 
It is unlikely that due to a few to ten percent uncertainties in
any group catalog, based on statistical errors and systematics,
one can use them directly to identify variations in clustering properties between groups of different ages or environments.
It is unclear if  the measurements in Yang et al. (2005b) can be
 interpreted as a conclusive test on the age dependence of the galaxy or halo bias. 
One can question why the same situation, or increase in bias, does not happen in the low mass bin, for e.g., between 
10$^{12}$ to $10^{12.5}$ M$_{\sun}$ as shown in Fig.~1. We find that if the contamination comes from adjacent mass bins, when $M < M_*$,
due to the fact that the halo bias is more or less constant at the low end (e.g., Mo \& White 1996; Seljak \& Warren  2004),
there is no significant difference in bias from the contaminants.
Small changes are more visible in the high mass end both due to small density of galaxies and the large changes to the halo bias
factor with a small change in the halo mass. It will be interesting to see if these variations will also show up in
similar tests with dark matter catalogs directly from numerical simulations whose measurements of dark matter properties will also
be affected by small uncertainties.

\section{Is there a more reliable test on the age dependence of galaxy properties?}

Instead of addressing whether bias itself is age dependent in the observed data, perhaps, 
it may be easier to address if galaxy properties of a given dark matter halo are defined only by the halo mass
 regardless of the age or the environment of the halo.  Analytically, the test is simple:
Is $\Phi(L|M,X) = \Phi(L|M)$, or is there a second parameter $X$
that determines galaxy properties of a halo in addition to the mass of that halo? This second parameter 
may be the environment, characterized by the overdensity $\delta_{\rm gal}$,
age at which the halo formed $t_f$, or another parameter that may 
determine galaxy formation and evolution (such as heating associated with reionization). 
It is now well known that the satellite LF of groups and clusters 
are different at the faint-end; one finds a larger fraction of dwarf 
galaxies in clusters while the faint-end LF of groups is flat. 
Whether the LF is different in clusters and groups when the age or
the environment of that group or cluster is varied 
may provide a useful test on the 
extent to which the assumption that the CLF depends only on mass is valid. 

Similarly, it may be useful to study differences in properties of ``fossil'' systems
relative to all other groups of the same mass. Fossil systems are expected to be older, isolated, and more 
biased relative to groups of the same mass. Perhaps the difference in properties between
fossil and all other groups capture the extent to which $\Phi(L|M,t_f) $ depends on
 the age of the halo $t_f$. Since the fraction of fossil systems at group mass scales are $\sim$ 10\% 
(Milosavljevi\'c et al. 2005), one should only expect variations in the galaxy sample roughly around the same level. 

One can easily address the issue on how large-scale environment affect galaxy properties by randomizing 
the galaxy distribution within halos and comparing statistics before and after the redistribution. This was considered by
Yoo et al. (2005) using a simulation and a semi-analytical model 
and found $\lesssim 5$\%  differences in clustering statistics involving the 
galaxy-galaxy and galaxy-mass cross-correlation functions. 
Whether these changes are a real reflection of galaxy properties and evidence for $\Phi(L|M,\delta_{\rm gal})$
rather than $\Phi(L|M)$ or a reflection of uncertainties in simulations and semi-analytical models is yet to be
seen. This and similar approaches only partly address the issue whether $\Phi(L|M,X)=\Phi(L|M)$ since
when one redistributes galaxies to different halos one cannot study differences coming from the
internal structure of that halo. For example, older halos may have a dark matter profile with a larger
concentration parameter and galaxy properties may reflect that compared to a same mass halo but formed later.
When one compares a redistributed sample to the real one, effects coming within halos will simply be averaged out.
Addressing whether  $\Phi(L|M,X)=\Phi(L|M)$ is challenging, but large catalogs of galaxies from surveys
such as SDSS and 2dF provide the opportunity for a variety of studies. Though how these catalogs will provide
the ultimate answer is yet to be seen.

One may be able to establish few percent differences with dependences in the environment through
a combination of luminosity functions and clustering as a function of overdensity.
These statistics depend on different combinations of $dn(\delta_{\rm gal})/dM$  and $\Phi(L|M,\delta_{\rm gal})$ (see, Cooray 2005c for expressions), for example, and the combination of one-point (luminosity function) and two-point (clustering statistics) 
allows degeneracies to be broken. Still, the test is challenging. Based on a simple Fisher matrix
test, using the concentration dependent bias and occupation properties of Wechsler et al. (2005)
as the basis for variations with age, we find that one is required to have 1\% accurate 
measurements in clustering statistics and
luminosity functions divided to $\sim$ 10 bins of some parameter that is taken to be a direct estimator of age.
The accuracy of establishing differences is limited because one must introduce variations in $b_{\rm halo}(M,\delta_{\rm gal}\; {\rm or} c)$
when describing clustering statistics, but not in the luminosity function. 

Finally, to conclude, there is a strong possibility that one must account for small (at most ten percent) changes
to galaxy properties with a second parameter in addition to the halo mass. This looks
challenging when compared to the simple halo model that has a built-in assumption that the 
halo occupation distribution is defined only by the halo mass.  If the second
parameter that vary properties of the conditional functions can be
established, this information can be easily included when interpreting data under an assumed cosmological model or when attempting to 
extract cosmology with galaxy clustering measurements. The conditional function approach, say compared to the
halo occupation distribution, allows this to be easily achieved since one can easily condition in terms of the mass
and the second parameter as well as accounting for any joint distribution between mass and that parameter.
We think that the conclusion that recent results invalidate analytical attempts to understand the galaxy distribution and use it for
cosmological parameter estimates is premature.

{\it Acknowledgments:} 
Authors acknowledges useful correspondences with Frank van den Bosch on results related to Yang et al. (2005b) and
useful discussions with James Bullock and Risa Wechsler on Wechsler et al. (2005).


\begin{thebibliography}{99}
\frenchspacing

\bibitem[Balogh et al. (2004)]{Bal04}
  Balogh, M. et al. 2004, MNRAS, 348, 1355

\bibitem[Blanton et al. (2004)]{Bla04}
  Blanton, M. R., Eisenstein, D. J., Hogg, D. W. \& Zehavi, I. 2004, astro-ph/0411037

\bibitem[Colless et al. (2001)]{Col01}
Colless, M. et al. 2001, MNRAS, 328, 1039

\bibitem[Cooray \& Sheth (2002)]{CooShe02}
  Cooray, A. \& Sheth, R. 2002, Physics Reports, 372, 1  (astro-ph/0206508)

\bibitem[Cooray \& Milosavljevi\'c(2005a)]{Cooray:05a}
Cooray, A., \& Milosavljevi\'c, M.\ 2005a, ApJ, 627, L85 (astro-ph/0503596)

\bibitem[Cooray \& Milosavljevi\'c(2005b)]{Cooray:05b}
Cooray, A., \& Milosavljevi\'c, M.\ 2005b, ApJ, 627, L89 (astro-ph/0504580)

\bibitem[Cooray (2005a)]{Coo05}
Cooray, A., 2005a, MNRAS, 363, 337 (astro-ph/0505421)

\bibitem[Cooray (2005b)]{Coo05b}
Cooray, A., 2005b, MNRAS, in press (astro-ph/0509033)

\bibitem[Dressler (1980)]{Dre80}
Dressler, A. 1980, ApJ, 236, 351

\bibitem[Dressler (1997)]{Dre97}
Dressler, A., Oemler, A. Jr., Couch, W. J., et al. 1997, ApJ, 490, 577

\bibitem[Gao et al. (2005)]{Gao05}
Gao, L., Springel, V. \& White, S. D. M. 2005, preprint (astro-ph/0506510)

\bibitem[Goto et al. (2003)]{Got03}
Goto, T. et al. 2003, MNRAS, 346, 601

\bibitem[Harker et al.(2005)]{har:05}
Harker, G., Cole, S., Helly, J., Frenk, C., \& Jenskins, A. 2005, preprint (astro-ph/0510481)

\bibitem[Jenskins et al. (2001)]{Jen01}
Jenkins, A. et al. 2001, MNRAS, 321, 372

\bibitem[Kravtsov et al. (2004)]{kra04}
  Kravtsov, A. V. et al. 2004, ApJ, 609, 35

\bibitem[Milosavljevi\'c et al. (2005)]{Milos05}
Milosavljevi\'c, M. et al. 2005, ApJL in press (astro-ph/0509647)

\bibitem[Mo \& White (1996)]{MoWhi96}
Mo, H. J., \& White, S. D. M. 1996, MNRAS, 282, 347

\bibitem[Mo et al. (2004)]{Mo04}
Mo, H. J. et al. 2004, MNRAS, 349, 205

\bibitem[Navarro et al. (1997)]{Navetal97}
Navarro, J. F., Frenk, C. S. \& White, S. D. M. 1997, ApJ, 490, 493

\bibitem[Seljak \& Warren (2004)]{Sel04}
Seljak, U. \& Warren, M. S. 2004, MNRAS, 355, 129

\bibitem[Sheth \& Tormen(1999)]{Sheth:99}
 Sheth, R.~K., \& Tormen, G.\ 1999, \mnras, 308, 119

\bibitem[Sheth, Mo \& Tormen(2001)]{Sheth:01}
 Sheth, R.~K., Mo, H. J., \& Tormen, G.\ 2001, \mnras, 323, 1

\bibitem[Vale \& Ostriker (2004)]{Val04}
  Vale, A. \& Ostriker, J. P. 2004, MNRAS, 353, 189

\bibitem[Wechsler et al. (2005)]{Wec05}
Wechsler, R. H. et al. 2005, preprint, astro-ph/0512xxx

\bibitem[Weinmann et al.(2005)]{Wein05}
Weinmann, S. M., van den Bosch, F. C., Yang, X. \& Mo, H. J. 2005, preprint (astro-ph/0509147)

\bibitem[Yang et al. (2003)]{Yan03}
Yang, X., Mo, H. J., \& van den Bosch, F. C. 2003, MNRAS, 339, 1057

\bibitem[Yang et al. (2005a)]{Yan05a}
Yang, X., Mo, H. J., Jing, Y. P., van den Bosch, F. C. 2005a, MNRAS, 358, 217

\bibitem[Yang et al. (2005b)]{Yan05b}
Yang, X., Mo, H. J., \& van den Bosch, F. C. 2005b, preprint (astro-ph/0509626)

\bibitem[Yoo et al. (2005)]{Yoo05}
Yoo, J.,  et al., preprint, astro-ph/0511580

\bibitem[Zheng \& Weinberg (2005)]{Zhe05}
Zheng, Z. \& Weinberg, D. 2005, preprint, astro-ph/0512071


\end{thebibliography}
\end{document}